\documentclass[aps,prl,superscriptaddress,showpacs,amsmath,twocolumn]{revtex4}

\usepackage{amsmath}
\usepackage{amssymb}
\usepackage{graphicx}

\begin{document}


\title{Peculiar Nature of Snake States in Graphene }

\author{ P. Rakyta 
}
\affiliation{Department of Physics of Complex Systems,
E{\"o}tv{\"o}s University, 
H-1117 Budapest, P\'azm\'any P{\'e}ter s{\'e}t\'any 1/A, Hungary
}

\author{L.~Oroszl\'any}
\affiliation{Department of Physics, Lancaster University,
Lancaster, LA1 4YB, UK}

\author{A.~Korm\'anyos}
\affiliation{Department of Physics, Lancaster University,
Lancaster, LA1 4YB, UK}

\author{C.~J. Lambert}
\affiliation{Department of Physics, Lancaster University,
Lancaster, LA1 4YB, UK}

\author{ J. Cserti
}
\affiliation{Department of Physics of Complex Systems,
E{\"o}tv{\"o}s University, 
H-1117 Budapest, P\'azm\'any P{\'e}ter s{\'e}t\'any 1/A, Hungary
}



\begin{abstract}

We study the dynamics of the electrons in a non-uniform magnetic field
applied perpendicular to a graphene sheet in the low energy limit
when the excitation states can be described by a Dirac type Hamiltonian. 
We show that as compared to the two-dimensional electron gas (2DEG) 
snake states in graphene exibit peculiar properties related to the
underlying dynamics of the Dirac fermions.  
The current carried by snake states is locally uncompensated 
even if the Fermi energy lies between the
first non-zero energy Landau levels of the conduction and valence bands. 
The nature of these states is studied by calculating the 
current density distribution.  
It is shown that besides the snake states in finite samples surface 
states also exist.

\end{abstract}


\maketitle


In recent experiments on graphene new transport phenomena resulting from
the massless Dirac fermion type excitations have been 
observed generating huge interest both experimentally
and theoretically~\cite{Novoselov-1:cikk,Kim:cikk,Novoselov-bilayer:cikk}.  
For reviews on graphene see
Refs.~\onlinecite{Katsnelson:rev,Katsnelson_Novoselov:rev,Geim_Novoselov:rev}
and a special issue in Ref.~\cite{Solid_State_Comm:cikkek}.

While for 2DEG a number of experimental 
and theoretical works have been devoted to study the excitation 
spectrum of electrons and their transport properties in non-uniform 
magnetic fields, only a little is known about graphene in this case.  
For example, in 2DEG a special state called snake state exists  
in non-uniform magnetic field at the boundary where the direction of 
the magnetic field changes. 
The snake states in 2DEG were first studied theoretically 
by M\"uller~\cite{Muller:cikk} and it inspired numerous 
theoretical and experimental works 
(see, eg, Ref.~\cite{non-uniB:cikkek} and references therein).  
The effect of an inhomogeneous magnetic field on electrons has been 
much less investigated in graphene. 
Martino et al.\ have demonstrated that the massless Dirac electron 
can be confined in inhomogeneous magnetic field~\cite{Martino:cikk}. 
Tahir and Sabeeh have studied the magneto-conductivity of graphene 
in a spatially modulated magnetic field and
shown that the amplitudes of the Weiss oscillation are larger than
that in 2DEG~\cite{Tahir:cikk}. 
The low energy electronic bands have been studied by Guinea et
al.~\cite{Guinea_corrug:cikk} taking into account the 
non-homogeneous effective gauge field due to the ripples of the graphene
sheet. 
However, snake states have been studied only in carbon nanotubes very
recently by Nemec and Cuniberti~\cite{Nemec:cikk}.

\begin{figure}[hbt]
\includegraphics[scale=0.7]{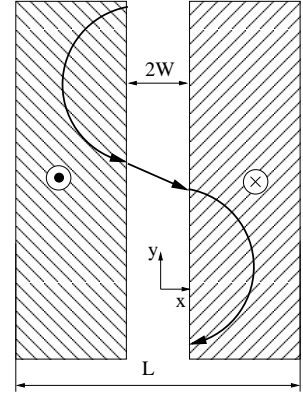}
\caption{\label{geo:fig} 
The magnetic field applied perpendicular to the graphene is zero 
in the center region of width $2W$, and at the two
sides of the strip they are in opposite directions with the same
magnitude $B$. 
The total width of the strip is $L \gg W$. 
The classical trajectory of a typical snake state is also drawn
schematically. 
}
\end{figure}

In this work we study the electronic properties of graphene in
a non-uniform magnetic field as shown in Fig.~\ref{geo:fig}. 
We show the existance of snake states exibiting peculiar behaviour at
low energy. 
In particular, we find that these snake states are localized 
in the zero magnetic field region and carry current which is 
compensated at the edges of the sample far from the central region.  

To study the nature of the snake states in graphene we consider 
a simple step like, non-uniform magnetic field applied perpendicular to
the graphene. 
The Dirac Hamiltonian of graphene in low energy approximation reads 
$H_{\pm} = v \left(\sigma_x \pi_x  \pm \sigma_y \pi_y \right)$, 
where the index $+$ ($-$) corresponds to the valley ${\bf K}$ 
(${\bf K}^\prime$), the Fermi velocity is $v=\sqrt{3}/2 \, a\gamma /\hbar$ 
given by the coupling constant $\gamma$ between the nearest neighbours 
in graphene ($a=0.246$ nm is the lattice constant in the honeycomb 
lattice), while 
$\mbox{\boldmath $\pi$} = (\pi_x,\pi_y) = {\bf p} - e {\bf A}$ is given
by the canonical momentum ${\bf p}$ and the vector potential ${\bf A}$
related to the magnetic field as ${\bf B}= \text{rot}{\bf A}$.
The Pauli matrices $\sigma_x$ and $\sigma_y$ 
act in the pseudo-spin space.
The Hamiltonian is valley degenerate for arbitrary magnetic field, ie, 
$\sigma_x H_{\pm} \sigma_x = H_{\mp}$,
therefore it is enough to consider only one valley. 
In what follows we take valley ${\bf K}$.

The energy spectrum of our system can be obtained from  
the Schr\"odinger equation $H_+ \Psi (x,y) = E\Psi (x,y)$. 
In our analytical calculations we assumed that $L \gg l$, where 
$l=\sqrt{\hbar /|eB|}$ is the magnetic lenght. 
To construct the wave function in each region we utilize 
the symmetries of the system. 
The system is translation invariant in the $y$ direction 
and $\left[H,p_y \right]=0$ in Landau gauge if  
the vector potential has a form ${\bf A}= (0,A_y(x),0){}^T$. 
Therefore the wave function can be separated: 
$\Psi (x,y) = \Phi (x) e^{i k y}$, where $k$ is the
wave number along the $y$ direction. 
One can also show that $\left[H,\sigma_y T_x\right]=0$, 
where $T_x$ is the reflection operator, ie, 
it acts on an arbitrary function $f(x)$ as $T_x f(x)=f(-x)$. 
This implies that the wave functions $\Phi(x)$ can be classified as even 
wave functions satisfying 
$\sigma_y T_x \Phi^{\left(e\right)}(x) = \Phi^{\left(e\right)}(x)$ 
and odd functions, when  
$\sigma_y T_x \Phi^{\left(o\right)}(x) = - \Phi^{\left(o\right)}(x)$. 
Moreover, it is true that $\left[\sigma_y T_x,p_y \right]=0$.   
From these commutation relations the even (odd) wave function
ansatz in the three different regions 
can be constructed.  
In the central region, ie, for $ |x| \le W$ and at energy $E$ 
it is given by 
\begin{subequations}
\begin{eqnarray}
\Phi_C^{\left(e\right)} (x) &=& 
c_1 \left[\left( \begin{array}{c}
1  \\ e^{i \varphi}  
\end{array}  \right) e^{i K x} -i   
\left( \begin{array}{c}
e^{i \varphi}  \\ -1   
\end{array}  \right) e^{-i K x}
\right], \\
\Phi_C^{\left(o\right)} (x) &=&  
c_1 \! \left[\left( \begin{array}{c}
1   \\  e^{i \varphi}  
\end{array}  \right) e^{i K x} + 
i \left( \begin{array}{c}
e^{i \varphi}  \\ -1     
\end{array}  \right) e^{-i K x}
\right] \!  ,
\end{eqnarray}%
where $\tan \varphi = k/K$, $K= \sqrt{\varepsilon^2 - k^2}$ is 
the transverse wave number and $\varepsilon = E/(\hbar v)$.   
In the left hand side, ie, for $ x \le - W $ the wave function in
Landau gauge  reads 
\begin{eqnarray}
\Phi_L^{\left(e\right)} (x) &=& 
c_2 \left( \begin{array}{c}
 U(a_+,\xi) \\[1ex] 
\eta\,  U(a_-,\xi)  
\end{array}  \right), \,\, 
\Phi_L^{\left(o\right)} =  \Phi_L^{\left(e\right)}, 
\end{eqnarray}%
\end{subequations}%
where $\xi = -\sqrt{2}\left(x+W + kl^2 \right)/l$,  
$a_{\pm}= \left(\pm 1-\varepsilon^2 l^2 \right)/2$,   
$\eta = -i \sqrt{2}/(\varepsilon l )$, and $U(a,x)$ is 
a parabolic cylinder function\cite{ref:abramowitz} (from the two
parabolic cylinder functions we take that which tends to zero 
for $x \to -\infty$).
The wave function ansatz in the right region, ie,  for $ x \ge W $ 
can be obtained from $\Phi_L^{\left(e,o\right)}(x)$ as 
$\Phi_R^{\left(e\right)}(x) = \sigma_y T_x \Phi_L^{\left(e\right)}(x)$
and 
$\Phi_R^{\left(o\right)}(x) =- \sigma_y T_x \Phi_L^{\left(o\right)}(x)$.   
When $k > \varepsilon $, the above given transverse wave number $K$ 
has to be replaced by $K= -i \sqrt{k^2-\varepsilon^2}$.  
The boundary conditions require the continuity of the
wave function at $x=\pm W$ resulting in a homogeneous system of
equations for the amplitudes $c_1$ and $c_2$. 
Hence, nontrivial solutions can be obtained from the secular 
equation resulting in energy bands $E_n(k)$ labelled by
$n=0,\pm 1, \dots$ for a given $k$.
Owing to the chiral symmetry ($\sigma_z H_{\pm} \sigma_z = - H_{\pm}$) 
we have $E_{-n}(k) = - E_n(k)$\cite{Ezawa:cikk}. 

\begin{figure}[hbt]
\hspace{-5mm}\includegraphics[scale=0.2]{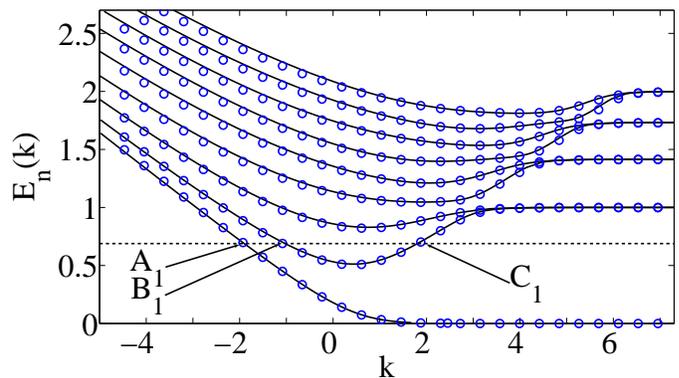}
\caption{\label{Ek:fig}  
The energy bands $E_n(k)$ (in units of $\hbar \omega_c$) 
as a function of $k$ (in units of $1/W$) around the ${\bf K}$ point 
for magnetic field corresponding to $W/l=2.2$.
The solid lines (open circles) are the results obtained from the Dirac
equation (TB model,~\cite{TB:note}). 
Only the conduction band ($E_n(k)\ge 0$) with $n=0, 1, \dots, 8$ states
is shown ($n=0$ corresponds to the lowest conduction subband, 
and the other bands are in increasing order in energy).
The even (odd) wave function states correspond to even (odd) quantum
number $n$. 
States $A_{1}$, $B_{1}$ and $C_{1}$ at energy $0.688\, \hbar \omega_c$
(dotted line) are snake states (see the text).
}
\end{figure}

\begin{figure}[hbt]
\includegraphics*[scale=0.35]{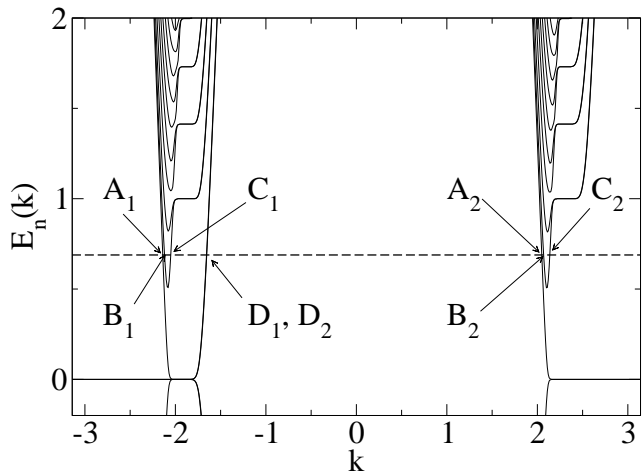}
\caption{\label{disp_gr_TB:fig}  
The dispersion relation $E_n(k)$ for all allowed $k$ (in units of $1/a$) 
including a small portion of the valence bands. 
The parameters are the same as in Fig.~\ref{Ek:fig} for TB calculation.   
At the same energy as in Fig.~\ref{Ek:fig} indicated by dashed line 
there are eight states:  
$A_i$, $B_i$, $C_i$ and $D_i$ with $i=1,2$ (see the text).
}
\end{figure}

Figure~\ref{Ek:fig} shows the comparison of energy bands $E_n(k)$ 
around the ${\bf K}$ point obtained from the Dirac equation 
and from tight binding (TB) model.  
Note that our TB calculation is similar to that made 
by Wakabayashi et al.\ except that they applied uniform magnetic 
field~\cite{Wakabayashi:cikk}. 
In the figure the surface states calculated from TB model are not shown. 
This will be discussed below. 
The energies are scaled in units of $\hbar \omega_c$, where  
$\hbar \omega_c= \sqrt{2} \hbar v /l = \sqrt{3/2}\, \gamma a/l $. 
One can see that the agreement between the two calculations is 
excellent. 
For large enough positive $k$ each state evolves
into dispersionless, twofold degenerate Landau levels    
having the same energies as in uniform magnetic field, ie, 
$E_n^L(k) = \text{sgn}(n)\, \hbar \omega_c\sqrt{|n|}$
with $n=0,\pm 1, \dots$, where $\text{sgn}(\cdot)$ is
the sign function (see, eg, Ref.~\cite{Dirac_LL:cikkek,Ezawa:cikk}). 
However, for negative wave number $k$ the energy bands are dispersive.
Examples for such states are $A_{1}$, $B_{1}$ and $C_{1}$ 
in Fig.~\ref{Ek:fig}.  
Each of these states carries current in the $y$ or $-y$ 
direction depending on their group velocity. 
The corresponding wave functions are localized in the central, 
zero magnetic field region as shown below (see Fig.~\ref{disp_gr_TB:fig}). 
These are the snake states in our system.

It is also clear from Fig.~\ref{Ek:fig} that the number of 
left and right moving states at a given energy are not the same 
which seems paradoxical at first sight. 
The ground state of the system appears to be unstable thermodynamically 
beacuse there is a net current flow in the $-y$ direction even in equilibrium. 
To understand this paradox one needs to consider the surface states
localized at the edges of the system ($x=\pm L/2$).
Figure~\ref{disp_gr_TB:fig} shows the same energy bands $E_n(k)$ as in
Fig.~\ref{Ek:fig} obtained from TB calculations for all allowed $k$. 
Two extra subbands appear and the resulting states at a given energy 
are denoted by $D_1$ and $D_2$ in Fig.~\ref{disp_gr_TB:fig} 
(although they are hardly distinguishable because of the parameters 
we used, see the caption of Fig.~\ref{Ek:fig}). 
We shall show that these non-degenerate states are surface states 
caused by the finiteness of the sample in the $x$ direction and 
carry current in the $y$ direction at the two edges, respectively.  
Moreover, in Fig.~\ref{disp_gr_TB:fig} states $A_{1,2}$, $B_{1,2}$, 
and $C_{1,2}$ are snake states and the first two states carry current 
in the $-y$, while $C_{1,2}$ in the $y$ direction.
For finite $L$ the surface states at the edges of the sample 
will compensate the net current carried by the snake states.  
It is easy to see that including the surface states 
the number of left and right moving states 
are the same for all energies ensuring a stable equilibrium state of 
the system. 
We expect the same result using the Dirac Hamiltonian 
for finite $L$ 
(see works by Brey and Fertig, and 
Abanin et al.\ in Ref.~\cite{Solid_State_Comm:cikkek}). 

\begin{figure}[hbt]
\includegraphics*[scale=0.8]{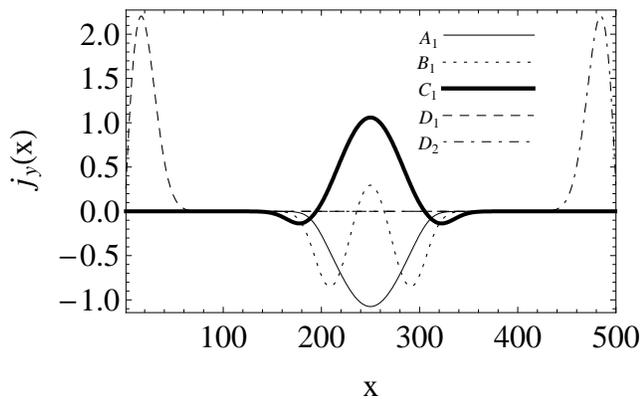}
\caption{\label{currents:fig}  
Current density distributions $j_y(x)$ (in arbitrary units) 
as a function of $x$ 
(in units of lattice site in $x$ ranging from $1$ to $N$)
for states $A_{1}$, $B_{1}$, $C_{1}$, $D_{1}$ and $D_{2}$ 
shown in Fig.~\ref{disp_gr_TB:fig}.  
}
\end{figure}

To get better insight into the nature of the snake and surface states
we calculated the current density distribution 
of these states shown in Fig.~\ref{currents:fig}. 
As can be seen from the figure the snake states $A_{1}$ and $B_{1}$ 
carry current in the $-y$ direction, while state $C_{1}$ 
in the opposite direction. 
Moreover, all of these states are localized in the center of the sample. 
States $D_{1}$ ($D_{2}$) are surface states with current 
flowing in the $y$ direction close to the left (right) edges of the sample. 
States $A_i$, $B_i$ and $C_i$ with $i=2$ behave the same way as
that with $i=1$ due to valley degeneracy. 

Varying the Fermi energy $E_F$ the character of these states would 
change and consequently their current distribution too. 
For Fermi energy lying between the first and second Landau 
levels, ie, for $E_0^L(k) < E_F < E_1^L(k)$ the net current
contributing from snake states $B_1$ and $C_1$ localized at the center of
the sample is zero, therefore they are locally compensated states.
While the snake state $A_1$ is also localized at the center, 
it is locally uncompensated.
Only the current from surface state $D_1$ will compensate 
the current from the snake state $A_1$ 
to ensure the stability of the ground state of the system.  
However, locally this snake state is uncompensated since the overlap
between states $A_1$ and $D_1$ is negligible. 
When the Fermi energy crosses the Landau level $n=1$ then not just state
$A_1$ but $B_1$ also becomes uncompensated, while state $C_1$ evolves
into a surface state. 
Thus, we find that in graphene for all Fermi energies 
there is always at least one locally uncompensated snake state localized at
the center of the sample and it is compensated only with surface
states far from the center part of the system.   

It is instructive to compare these results with that obtained for 
2DEG 
with the same magnetic field profile as in Fig~\ref{geo:fig}. 
\begin{figure}[hbt]
\includegraphics[scale=0.73]{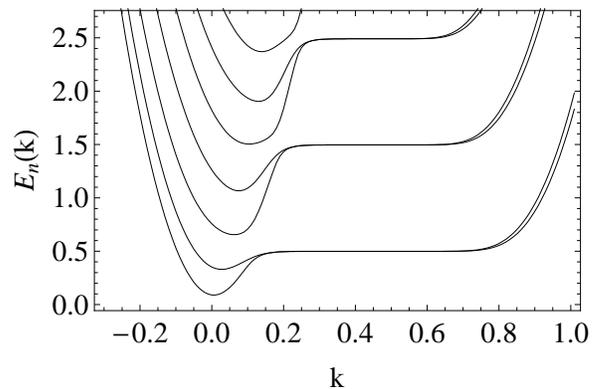}
\caption{\label{Ek_square:fig}  
The energy bands $E_n(k)$ of 2DEG obtained from TB model for the same 
magnetic field profile as in Fig.~\ref{geo:fig} and with magnetic
strength as in Fig.~\ref{Ek:fig}. 
The energies are in units  of $\hbar \Omega_c $, where
$\Omega_c=eB/m$ with effective mass $m$ of electrons, and 
$k$ is in units of $1/a$, where $a$ is the 
lattice constant in the square lattice). 
Here $W = L/10$ and $L= 200\, a$.
}
\end{figure}
As can be seen from Fig.~\ref{Ek_square:fig} the Landau levels are 
at energies $E_n(k)=\hbar \Omega_c (n+1/2)$, where $n=0,1,
\dots$ as expected in case of uniform magnetic field. 
Using a similar analysis as before for graphene 
one can find snake and surface states~\cite{sq_snake:note}.
The dispersion relation implies that in 2DEG no locally uncompensated 
current appears below the first Landau level ($n=0$).
This is a striking difference between the two systems.

In summary, we studied the dynamics of electrons in graphene when the
applied magnetic field is non-uniform. 
For a simple, step-like magnetic field dependence we show that snake
states appear similarly to the case of 2DEG.
We calculated the energy bands in case of an infinite system 
using the Dirac Hamiltonian which agree very well with that obtained 
from our TB calculations. 
Moreover, we show that the surface states of a finite width sample 
ensure the stability of the ground state of the system. 
We find that in contrast to 2DEG the snake
states in graphene  are locally uncompensated for all Fermi energies.
The different behavior of the snake states in graphene compared to
the 2DEG is a clear manifestation of the massless
Dirac fermion like excitations in graphene. 
The current carrying snake state at low enough Fermi energy (between
the Landau level $n=0$ and $n=1$ or  $n=-1$) is expected to be as 
resilient against scattering on impurities as the surface states 
in quantum Hall effect of graphene. 
The peculiar behavior of the snake states in
graphene with non-uniform magnetic fields could be utilized in future
experimental and theoretical works. 

We gratefully acknowledge discussions
with E. McCann, V. Fal'ko, F. Guinea and H. Schomerus.
This work is supported partially 
by European Commission Contract No.~MRTN-CT-2003-504574 and EPSRC.


\begin{thebibliography}{10}
\bibitem{TB:note}
Only nearest neighbor hopping was taken into account 
with zig-zag edges at $x=\pm L/2$.
Our results are unaffected by other types of edge at $x=\pm L/2$ 
when $L\gg W$.  
The magnetic field is taken into account by the usual Peirels
substitution. 
In our numerical calculations we take $W = L/10$ and 
$N= 500$ lattice sites in $x$ direction, thus $L=(\sqrt{3}N/2-1)a$.
\bibitem{sq_snake:note}
Regarding the snake states similar results were obtained 
in earlier studies see, eg, Reijniers and Peeters work 
in Ref.~\cite{non-uniB:cikkek}.
\end{thebibliography}
\end{document}